# Preparation of silver nanopatterns on DNA templates

Shuxi Dai[a,b], Xingtang Zhang[b], Tianfeng Li[b], Zuliang Du[b,*], Hongxin Dang[a,b]

[a]*State Key Laboratory of Solid lubrication, Lanzhou Institute of Chemical Physics,
Chinese Academy of Sciences, Lanzhou 730000, PR China*
[b]*Key Laboratory of Special Functional Materials, Henan University, Kaifeng 475001, PR China*

**Abstract**

Patterns of silver metal were prepared on DNA networks by a template-directed selective deposition and subsequent metallization process. Scanning force microscopic observations and XPS investigations demonstrated that uniform networks of nanosized silver metal clusters formed after incubation of DNA LB films with silver ions and subsequent chemical reduction of silver ions/DNA films samples. The results showed that this template-directed metallization on DNA LB films provided a simple and effective method for the construction of functional nanocomposite films.

## 1. Introduction

The special molecular structure of DNA has been shown to be useful in developing novel nanostructured materials [1–4]. By using DNA as building blocks for the programmed assembly of nanoscale structures, the assembly of two-dimensional DNA crystals, the assembly of gold nanoclusters using DNA as the linking molecules have been reported [1,2]. Possible applications of DNA molecules in electronic devices were inspired by the special electronic properties of DNA molecules [3–5]. DNA has a high affinity for metal cations, and these localized cations can be reduced to form metallic nanoparticles that follow the contour of the DNA template. In previous reports, metal nanowires on the DNA template were generated by the initial binding of $Ag^+$ ions to the DNA, followed by reduction of the ions to yield catalytic sites for the cluster growth [4]. By this method, palladium and platinum clusters grow on DNA molecule templates to fabricate palladium and platinum nanowires [5,6]. The conductivity results have demonstrated that wire-like metallic structures can be used directly in the electronic circuits [4–6].

Recently many research groups are attempting to fabricate DNA film, network and other patterned constructions and various methods have been used such as DNA molecular deposition, immobilization of

* Corresponding author. Tel.: +86 378 2193262;
fax: +86 378 2867282.
*E-mail address:* zld@henu.edu.cn (Z. Du).

DNA on SAMs [7,8]. In our previous studies, interfacial properties and organization process of DNA/ODA mixed Langmuir monolayers at the air/water interface were investigated [9]. Large-scale DNA patterns were fabricated through the combination of DNA and cationic surfactants by LB technique. In this work, we describe a new approach that use patterned DNA LB films with fractal surface as two-dimensional templates for the fabrication of nanocomposite films. Nanopatterns of silver metal were fabricated by chemical deposition of silver metal onto DNA LB film. To gain further information on the structure and chemical state of the nanocomposite films, we describe the complementary use of dynamic force microscope (DFM) and X-ray photoelectron spectroscopy (XPS) to characterize the immobilization of DNA templates and subsequent metallization of silver composite films. We aim to construct large-scale patterns of metal and semiconductor by using DNA networks as templates, and use the nanocomposite films in the application of nanoelectronic circuits.

## 2. Experiment

### 2.1. Materials

Calf thymus deoxyribonucleic acid (DNA) and octadecylamine (ODA) were purchased from Sigma (St. Louis, MO) and used without further purification. De-ionized water for all experiments was purified to a resistance of 18 M$\Omega$ cm. The DNA samples were diluted with de-ionized water to concentrations of 3 $\mu$g/ml (pH = 6.5). ODA was dissolved in chloroform (A.R. grade) at a concentration of 1 mM to form monolayers on the subphase. Silver nitrate (Aldrich, 99.998%) and hydroquinone (Sigma) were used as received.

### 2.2. Preparation of DNA LB films

The $\pi$–$A$ isotherms and Langmuir–Blodgett films transfer experiments were performed on ATEMETA LB 105 trough (France). The DNA/ODA complex monolayers were obtained by spreading the chloroform solutions of ODA on the aqueous DNA solutions with concentration of 3 $\mu$g/ml using microsyringe. The solvent was allowed to evaporate for 20 min before the isotherms were measured. Following compression to the desired surface pressure, the mixed monolayers of ODA/DNA were transferred to freshly cleaved mica substrates using a vertical dipping method to form DNA LB films. The whole process of formation of DNA LB films templates includes the adsorption, surface diffusion, nucleation and aggregation process of DNA molecules.

### 2.3. Preparation of silver nanopatterns on DNA LB films

The metallization process of DNA molecules was based on the reported procedure [4]. The DNA LB

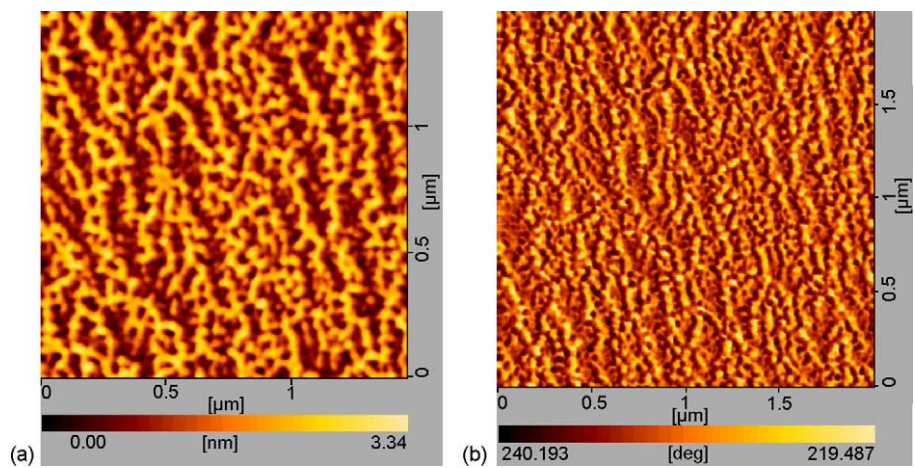

Fig. 1. Topography (a) and phase (b) images of 2000 nm × 2000 nm of DNA LB film on mica transferred under 20 mN/m.

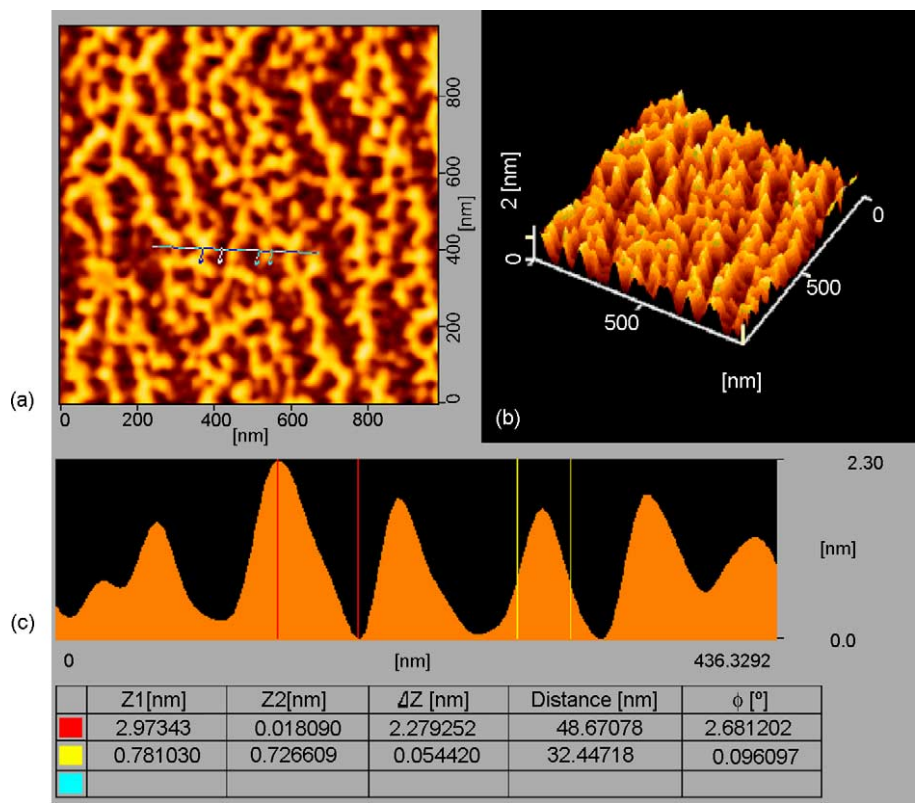

Fig. 2. Topography image of DNA LB film on mica of 1000 nm × 1000 nm (a) height image, (b) 3D image, (c) line profile.

films were placed into a 0.1 M $AgNO_3$ aqueous solution (ammonium hydroxide, pH = 10.5) to allow the selective localization of silver ions along the DNA molecules. After 10 min, the silver ion/DNA films was reduced by dipping the films in an ammonium hydroxide (pH = 10.5) medium containing 0.05 M hydroquinone to form small silver aggregates bound to the DNA molecules. Finally, an acidic solution (pH = 3.5, citrate buffer) of hydroquinone (0.05 M) and silver ions (0.1 M) was added in the dark to develop cluster growth. The samples were rinsed carefully afterwards to avoid homogeneously grown clusters to settle down on the films.

### 2.4. Atomic force microscopy (AFM)

AFM images of all samples were obtained in air with a commercial SPA400 SPM instrument (Seiko Instruments, Japan). A dynamic force microscope (DFM) mode with phase imaging was performed to avoid damage to the sample surface by the tip while scanning. Phase imaging can provide us additional information in detecting variations in compositions, adhesions and viscoelasticity properties of blended (especially polymer) samples [10,11]. The scanner size was 20 μm × 20 μm, and 100 μm long cantilevers with a spring constant of 20 N/m, resonance frequency 120 KHz and $Si_3N_4$ tips (Nanosensor Co.) were used. The temperature of the AFM experiments was 24 °C.

### 2.5. X-ray photoelectron spectroscopy (XPS)

X-ray photoelectron spectroscopy (XPS) measurements were done using an Axis-Ultra X-ray photoelectron spectrometer from Kratos (UK). Spectra were obtained at 90° takeoff angles using monochromatic Al Kα X-rays at 150 W (15 kV, 10 mA). Pass energies of 80 and 40 eV were used for survey and high-resolution scans, respectively. Postprocessing of the high-resolution XPS spectra for peak fitting and display was done off-line using Vision Processing analysis software (Kratos, UK) running on the Sun workstation.

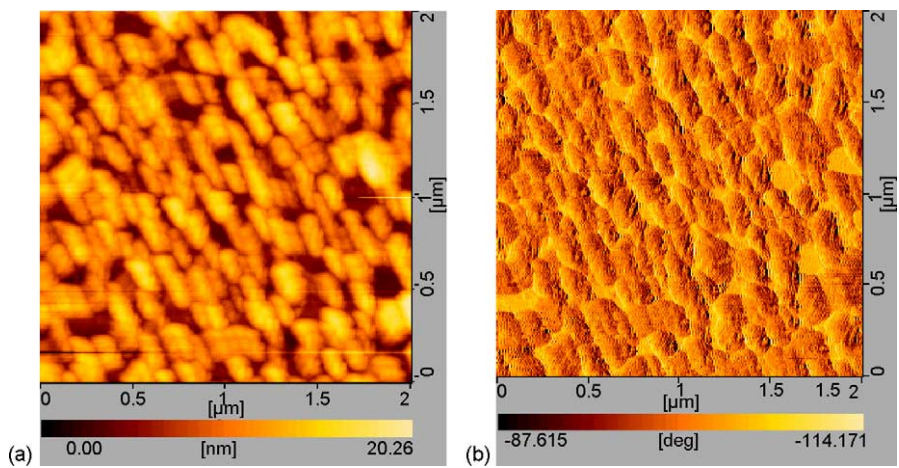

Fig. 3. Topography (a) and phase image (b) of 2000 nm × 2000 nm of silver metal patterns based on DNA templates.

## 3. Results and discussion

Fig. 1 shows typical 2000 nm × 2000 nm DFM images (topography and phase) of DNA LB film transferred onto mica under surface pressure of 20 mN/m. In our DFM measurements, the bright features in the topography images correspond to higher height in the films. The topography image (Fig. 1a) shows that mica substrates are covered by two-dimensional bright fibers and interconnected networks. The phase image (Fig. 1b) illustrates fractal patterns formed in the films and distributed over the surface quite uniformly. We attribute the interconnected fibers to aggregates of DNA molecules from the reported similar morphology of immobilized DNA on SAMs [7,8]. Individual DNA fibers are linking together to form nice patterns with high density of surface coating. Single DNA fiber can be resolved with different lengths ranging from 100 to 150 nm in the 1000 nm × 1000 nm topography image of Fig. 2. We measure the height of the DNA fibers and get an average height for 2.3 nm from the line profile analysis (Fig. 2c), which is near to value of theoretical diameter of ds-DNA molecules (2 nm) [12,13]. It indicates that DNA fibers are mainly formed as a monolayer of DNA molecules, and the DNA strands are parallel intertwined to the substrate surface. According to different AFM tips and imaging environment, the widths of calf thymus DNA strands for DNA immobilized on SAMs had been obtained with values ranging from 10 to 20 nm in many previous reports [12–14]. The widths of DNA fibers in Fig. 2 have an average width of ca. 40 nm, it indicates that several DNA molecules aggregated paralleled to the substrate surface to form a fiber seen in the AFM images. And then DNA fibers linked together to form high densities of DNA network coating on mica.

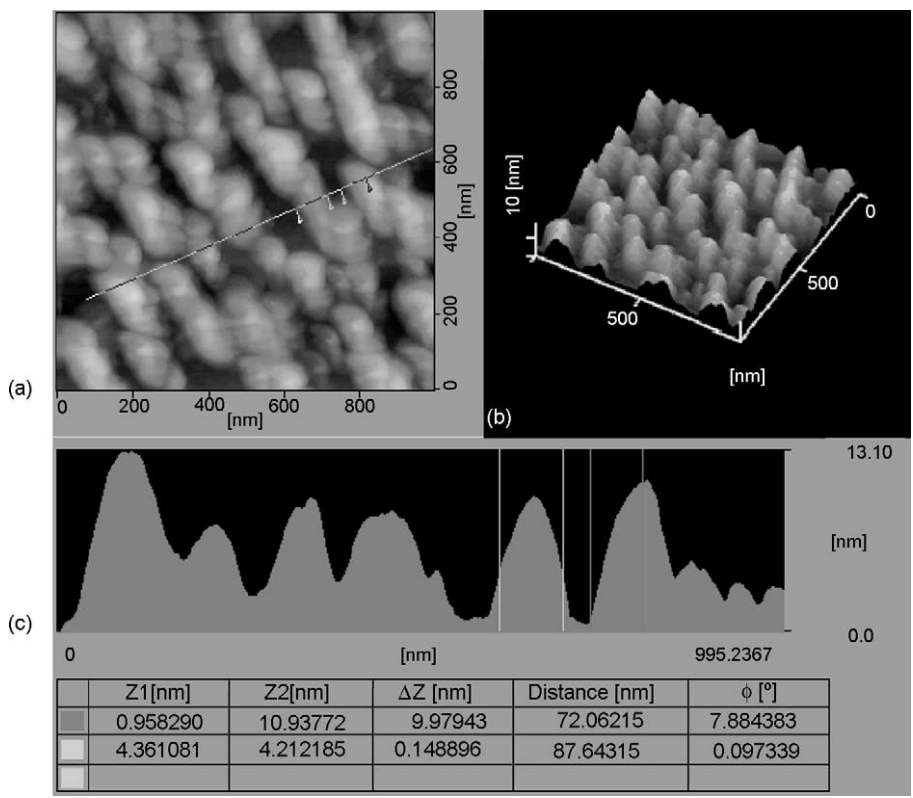

Fig. 4. Topography image of silver metal patterns on mica of 1000 nm × 1000 nm (a) height image, (b) 3D image, (c) line profile.

DFM images (topography and phase, 2000 nm × 2000 nm) of thin silver film chemical deposited on DNA LB film are shown in Fig. 3. The height image (Fig. 3a) shows that the surface was covered by two-dimensional nanoparticles networks with uniform arranged structure. The phase image (Fig. 3b) also illustrates the particles form nice surface pattern and distribute over the surface quite uniformly. A height analysis was performed on 1000 nm × 1000 nm topography image in Fig. 4. The images showed that silver nanoparticles coated on DNA LB films templates and remained the patterned surface after the metallization process. From the cross-sections analysis of Fig. 4c we obtain the average width of 80 nm and height of 10 nm for a single silver-coated nanofiber, which are much larger than those data obtained form DNA fibers in the LB templates. It can be seen clearly that silver metal deposition occurs only on the patterned DNA fibers.

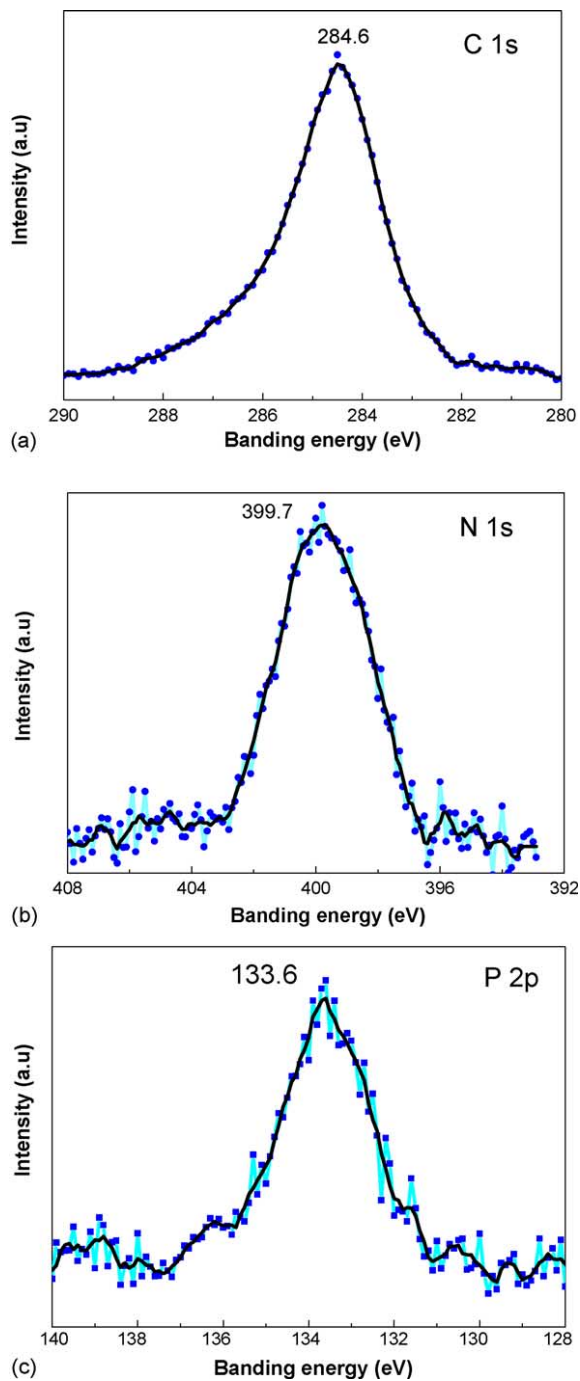

Fig. 5. XPS spectra of C 1s, N 1s and P 2p core-level recorded from the DNA LB films. (Filled circles for raw data, thick lines for total fits.)

To provide further information on the structure and chemical state of DNA LB films and silver composite films, we obtained the XPS spectra of the samples. A chemical analysis of DNA LB films was done using XPS and the C 1s, N 1s and P 2p core-level spectra were recorded. The spectra obtained are shown in panels of a–d, respectively, of Fig. 5. The C 1s spectrum showed the presence of a single component at 284.6 eV and is assigned to the carbons of ODA surfactants and DNA molecules. The N 1s spectrum shows the presence of a single component at 399.7 eV and it is assigned to the nitrogens in the bases of the DNA molecules in the LB films as well as the nitrogens in the ODA molecules [14]. A P 2p signal was recorded from the DNA LB film (Fig. 5c) and it was a good indicator for the presence of DNA molecules in the films. The p 2p BE agrees fairly well with reported values of DNA films immobilized on self-assembled monolayer surfaces and indicates no degradation of the DNA molecules [14,15].

We also measured the P 2p and Ag 3d core-level of silver metal patterning after the chemical deposition of silver nanoparticles. XPS spectra for C 1s, N 1s, P 2p and Ag 3d for silver metal films are showed in Fig. 6. The positions and intensities for P 2p and Ag 3d are compared for DNA LB films and silver metal patterning. It is clear that the Ag 3d peaks are present and the P 2p peaks are absent. The Ag $3d_{3/2}$ and Ag $3d_{5/2}$ peaks are identified at 374.2 and 368.2 eV, respectively. The XPS signals establish that the particles patterned on DNA LB film are primarily silver metal ($Ag^0$). The primary information obtained from XPS is the atomic concentrations of elements in the topmost 10 nm of the surface. From comparison of the AFM images of DNA LB films and silver metal films, we can conclude that the net heights of silver coating along DNA molecules is about 5–7 nm, But the P 2p signal of silver patterned films is too weak to be resolved, which indicates that the backbone of immobilized DNA molecules are mainly coated by silver nanoparticles and the silver nanoparticles grow just along the DNA networks of the DNA LB films templates.

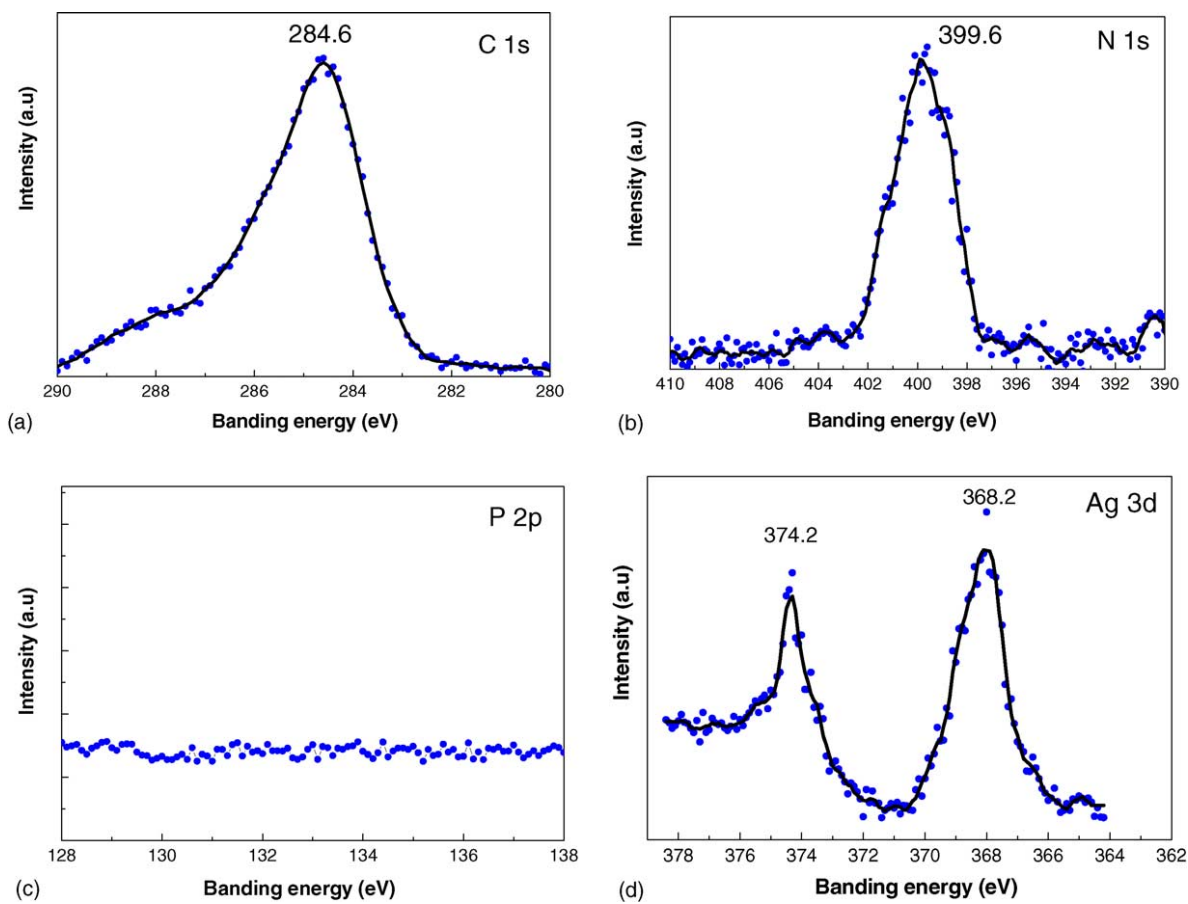

Fig. 6. XPS spectra of C 1s, N 1s, P 2p and Ag 3d core-level recorded from the silver metal patterns based on DNA LB film templates. (Filled circles for raw data, thick lines for total fits.)

From above AFM and XPS measurements, we demonstrate that silver metal patterns were prepared on DNA LB film templates by subsequent metallization process. The results of metallization experiments show that DNA networks act as very efficient nucleation centers for the growth of small crystalline particles. The silver nanoparticles only grow along the interconnected DNA fibers to form silver metal patterns. Under our experimental condition, silver ions strongly favor association with the heterocyclic bases and not the phosphates of the DNA [4]. The complexation of silver ions with bases can be demonstrated by the changes in the electronic absorption and circular dichroism spectra [16]. Firstly, incubation of DNA LB films in the silver nitrate aqueous solution made the silver ions selectively localize along the DNA fibers in the LB films and form complexes with the DNA bases. The silver ions/DNA films were then reduced to form metallic silver nanoparticles bound to the DNA fibers. Silver metal patterns on the substrates were formed with the template-direct effect of DNA networks. The size of the metal nanoparticles can be controlled by adjusting the concentration of silver ions and the developing time. The results also suggest that it may be feasible to control the formation of the nanopatterns of silver metal by changing the DNA LB templates and conditions of chemical reduction. The investigation of size-dependent optical and electronic properties of prepared silver nanopatterns is underway.

## 4. Conclusions

In summary, patterns of silver metal were prepared using DNA LB films templates by subsequent metallization process. XPS and AFM investigations showed that silver metal patterns were chemical deposited along the DNA networks and coated the DNA patterns of templates with silver nanoparticles to form uniform silver nanopatterns. We have demonstrated in detail a procedure for the selective growth of metal on DNA templates according to a template-direct nucleation and growth mechanism. It is shown that the prepared DNA templates can be used for selective deposition and metallization to construct large-scale metal composite films. The described metallization method is simple and suitable to be extended to larger scale processing and other inorganic materials. The composite nanoscale patterns of metal or semiconductor prepared based on DNA templates would be promising and ideal materials in the applications of nanoscale electronics.

## Acknowledgements

This work was supported by Natural Science Foundation of China (Nos. 90306010 and 20371015) and State Key Basic Research "973" Plan of China (No. 2002CCC02700).